\documentclass[pra,showpacs,nobalancelastpage]{revtex4}
\usepackage{graphicx}
\usepackage{amsmath}
\usepackage{amsfonts}
\usepackage{amssymb}

\begin{document}

\title{Generating continuous variable quantum codewords in the near-field atomic lithography}
\author{Stefano Pirandola, Stefano Mancini, David Vitali, and Paolo Tombesi}
\affiliation{ INFM, Dipartimento di Fisica, Universit\`a di
Camerino, via Madonna delle Carceri, I-62032 Camerino, Italy}

\date{\today}
\begin{abstract}
Recently, D. Gottesman \textit{et al.} [Phys. Rev. A \textbf{64},
012310 (2001)] showed how to encode a qubit into a continuous
variable quantum system. This encoding was realized by using
non-normalizable quantum codewords, which therefore can only be
approximated in any real physical setup. Here we show how a
neutral atom, falling through an optical cavity and interacting
with a single mode of the intracavity electromagnetic field, can
be used to safely encode a qubit into its external degrees of
freedom. In fact, the localization induced by a homodyne detection
of the cavity field is able to project the near-field atomic
motional state into an approximate quantum codeword. The
performance of this encoding process is then analyzed by
evaluating the intrinsic errors induced in the recovery process by
the approximated form of the generated codeword.
\end{abstract}
 \maketitle

\section{Introduction}

During last years quantum information and computation have been
extended to the continuous variable (CV) framework \cite{cvbook}.
In this framework, quantum and classical information is encoded
and processed using quantum systems, like oscillators and
particles, which are described by observables with a continuous
spectrum of eigenvalues. Recently, also quantum error correction
(QEC) has been extended to this framework, in order to allow a
reliable CV quantum computation \cite{Lloyd2,preskill}. In fact,
as shown in Ref.~\cite{preskill}, a whole class of CV QEC codes
can be designed to fight the effects of decoherence over a set of
particles, the most probable effect being a small diffusion in the
position and momentum of all the particles. These codes have been
suitably derived by extending to CV systems the
\textit{shift-resistant} quantum codes for qudits, and they can be
used to implement an universal set of fault-tolerant quantum gates
\cite{preskill}. However, the main drawback of the CV QEC seems to
be the physical preparation of the quantum codewords. In fact,
these codewords ideally are non-normalizable states (since
superpositions of infinitely squeezed states) and, in any real
physical implementation, they can only be approximated by
normalizable states, which will consequently introduce intrinsic
errors in the recovery process. Some literature has been devoted
to the development of schemes and techniques able to reduce the
intrinsic error probability in the recovery process. In
particular,\ Ref.~\cite{Travaglione0} has resorted to a sequence
of operations similar to a quantum random walk algorithm
\cite{Travaglione}. More recently, we have proposed an all-optical
scheme, based on the cross-Kerr interaction \cite{epl}, and a
trapped ion scheme \cite{ponde}, based on a ponderomotive
interaction \cite{Giovannetti}.

Here, revisiting some results of Refs.~\cite{walls}
and~\cite{wallsB} on atomic lithography, we show how to embed a
qubit in the external degrees of freedom of a free neutral atom.
As in Ref.~\cite{walls}, we consider a two-level atom passing
through an optical cavity and interacting with a single mode of
the intracavity electromagnetic field. Then, by making a
quadrature phase measurement on the field, it is possible to
localize the position of the atom within the wavelength of the
light in the cavity which acts as a virtual diffraction grating.
Depending on the initial state, the field measurement may localize
the atomic position wavefunction into one or more virtual slits.
In particular, we are interested in the case of an input state
sufficiently delocalized in position, so that the
measurement-induced localization of the atom will result in the
generation of a comb-like state, which represents the
finite-energy approximation of an ideal CV quantum codeword.

The paper is organized as follows. In Sec.~\ref{ideal} we rapidly review some
elements from Ref.~\cite{preskill} and we turn from an ideal situation to a
more realistic one. In Sec.~\ref{fly_scheme} the physical implementation of
the encoding scheme is proposed. Sec.~\ref{conclusion} is for conclusions.

\section{Continuous variable quantum codewords\label{ideal}}

\begin{figure}[ptbh]
\vspace{-0.0cm}
\par
\begin{center}
\includegraphics[width=0.6\textwidth] {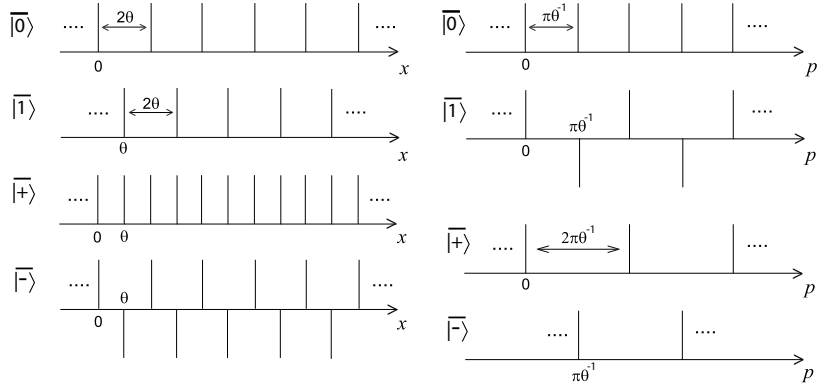}
\end{center}
\par
\vspace{-0.3cm}\caption{Ideal encoded states $\overline{\left|  0\right\rangle
},\overline{\left|  1\right\rangle }$ ($\bar{Z}$ eigenstates) and
$\overline{\left|  +\right\rangle },\overline{\left|  -\right\rangle }$
($\bar{X}$ eigenstates). On the left the structure of the spatial
wavefunctions is displayed while on the right the structure of the momentum
wavefunction is displayed. Each spike is ideally a Dirac-delta function.}%
\label{fig00}%
\end{figure}

A single qubit living in a Hilbert space $\mathcal{H}$ with basis $\{\left|
0\right\rangle ,\,\left|  1\right\rangle \}$ can be encoded into a single
particle \cite{particle} in such a way that the two resulting codewords
$\overline{\left|  0\right\rangle },\overline{\left|  1\right\rangle }$
provide protection against small diffusion errors in both position $x$ and
momentum $p$ (the quantum operators obey the commutation rule $[\hat{x}%
,\hat{p}]=i$ so that $x,p$ are dimensionless quantities). The two quantum
codewords $\overline{\left|  0\right\rangle },\overline{\left|  1\right\rangle
}$ are the simultaneous eigenstates, with eigenvalue $+1$ of the displacement
operators $\hat{D}_{x}(2\theta),$ $\hat{D}_{p}(2\pi\theta^{-1})$ with
$\theta\in\mathbb{R}$, which are also the stabilizer generators of the code
\cite{stabilizer}. These codewords are therefore invariant under the shifts
$x\rightarrow x-2\theta$ and $p\rightarrow p-2\pi\theta^{-1}$. Up to a
normalization factor they are given by
\begin{align}
\overline{\left|  0\right\rangle }  &  =\sum_{s=-\infty}^{+\infty}\left|
x=2\theta s\right\rangle =\sum_{s=-\infty}^{+\infty}\left|  p=\pi\theta
^{-1}s\right\rangle \label{ideal_0}\\
\overline{\left|  1\right\rangle }  &  =\sum_{s=-\infty}^{+\infty}\left|
x=2\theta s+\theta\right\rangle =\sum_{s=-\infty}^{+\infty}(-1)^{s}\left|
p=\pi\theta^{-1}s\right\rangle =\hat{D}_{x}(\theta)\overline{\left|
0\right\rangle } \label{ideal_1}%
\end{align}
i.e. they are a coherent superposition of infinitely squeezed states (position
eigenstates and momentum eigenstates). Each of them is a comb-state both in
$x$ and in $p$ with equally spaced spikes ($2\theta$ in $x $ and $\pi
\theta^{-1}$ in $p$). The codewords $\overline{\left|  0\right\rangle
},\overline{\left|  1\right\rangle }$ are also eigenstates of the encoded
bit-flip operator $\bar{Z}=\hat{D}_{p}(\pi\theta^{-1})$. Equivalently one can
also choose the codewords $\overline{\left|  \pm\right\rangle }=[\overline
{\left|  0\right\rangle }\pm\overline{\left|  1\right\rangle }]/\sqrt{2}$
which are the eigenstates of the encoded phase-flip operator $\bar{X}=\hat
{D}_{x}(\theta)$ and are given by:
\begin{align}
\overline{\left|  +\right\rangle }  &  =\sum_{s=-\infty}^{+\infty}\left|
x=\theta s\right\rangle =\sum_{s=-\infty}^{+\infty}\left|  p=2\pi\theta
^{-1}s\right\rangle \label{ideal_piu}\\
\overline{\left|  -\right\rangle }  &  =\sum_{s=-\infty}^{+\infty}%
(-1)^{s}\left|  x=\theta s\right\rangle =\sum_{s=-\infty}^{+\infty}\left|
p=2\pi\theta^{-1}s+\pi\theta^{-1}\right\rangle . \label{ideal_meno}%
\end{align}
Also these states are comb-like states both in $x$ and in $p$, with equally
spaced spikes ($\theta$ in $x$ and $2\pi\theta^{-1}$ in $p$). The four
codewords states are schematically displayed in Fig.~\ref{fig00}.

The recovery process is realized by measuring the stabilizer generators
$\hat{D}_{x}(2\theta),$ $\hat{D}_{p}(2\pi\theta^{-1})$. The measurement of the
$X$-generator $\hat{D}_{x}(2\theta)=e^{-i2\theta\hat{p}} =\hat{p}($%
mod$\pi\theta^{-1})$ reveals momentum shifts $\Delta p$ which are correctable
if $\left|  \Delta p\right|  <\pi\theta^{-1}/2$; in such a case the correction
is made by shifting $p$ so to become equal to the nearest multiple of
$\pi\theta^{-1}$. In the same way, the measurement of the $Z$-generator
$\hat{D}_{p}(2\pi\theta^{-1})=e^{i2\pi\theta^{-1}\hat{x}}=\hat{x}($%
mod$\theta)$ reveals position shifts which are correctable if $\left|  \Delta
x\right|  <\theta/2$; in such a case the correction is made by shifting $x$ so
to coincide with the nearest multiple of $\theta$.

Ref.~\cite{preskill} proposed the following recipe for the generation of the
codeword states.

\begin{enumerate}
\item  Preparation of a particle in the $p=0$ eigenstate (i.e. completely
delocalized in position).

\item  Coupling the particle to a meter (i.e. an oscillator, with ladder
operators $\hat{c}$, $\hat{c}^{\dagger}$) via the non linear interaction
$\hat{H}_{NL}=g\hat{c}^{\dagger}\hat{c}\hat{x}$. This interaction modifies the
frequency of the meter by $\Delta\omega=gx$ so that, at time $t=\pi\theta
^{-1}g^{-1}$, the phase of the meter is shifted by $\Delta\phi=\pi\theta
^{-1}x$.

\item  Reading out the phase of the meter $\Delta\phi$ at a time $t$, i.e.
measuring $\hat{x}($mod$2\theta)$. This measurement projects the initial state
into a superposition of equally spaced delta function $\delta(x-2\theta
s+\varepsilon)$ with $s=0,\pm1,...$ and $\varepsilon\in\mathbb{R}$.

\item  Applying a suitable transformation to obtain any desired encoded qubit
state $a\overline{\left|  0\right\rangle }+b\overline{\left|  1\right\rangle
}$.
\end{enumerate}

Ideally the codewords are non-normalizable states infinitely
squeezed both in $x$ and $p$, but in practice one can only
generate states with finite squeezing, i.e. only approximate
codewords: $\widetilde{\left|  0\right\rangle
}\sim\overline{\left|  0\right\rangle },$ $\widetilde{\left|
1\right\rangle }\equiv\hat{D}_{x}(\theta)\widetilde{\left|
0\right\rangle }\sim \overline{\left|  1\right\rangle },$
$\widetilde{\left|  \pm\right\rangle
}\equiv\lbrack\widetilde{\left|  0\right\rangle
}\pm\widetilde{\left| 1\right\rangle
}]/\mathcal{N}_{\pm}\sim\overline{\left|  \pm\right\rangle }$
($\mathcal{N}_{\pm}$ are normalization constants). For this
reason, in order to estimate the quality of the encoding scheme,
together with the error probability in the recovery process due to
the occurrence of an uncorrectable error, we have also to consider
the \textit{intrinsic error probability} due to the imperfections
of the approximate codewords which can lead to an error even in
the presence of a correctable error. Here we propose a physical
implementation of the ideal coding protocol of Ref.~
\cite{preskill} based on single neutral atoms interacting with a
radiation mode. It can be derived from the ideal one by replacing
the initial $p=0$ state with a finitely squeezed state,
$\hat{H}_{NL}$ with a ponderomotive interaction, and the phase
measurement with a homodyne measurement.

\section{Encoding by atomic lithography\label{fly_scheme}}

\begin{figure}[ptbh]
\begin{center}
\includegraphics[width=0.40\textwidth] {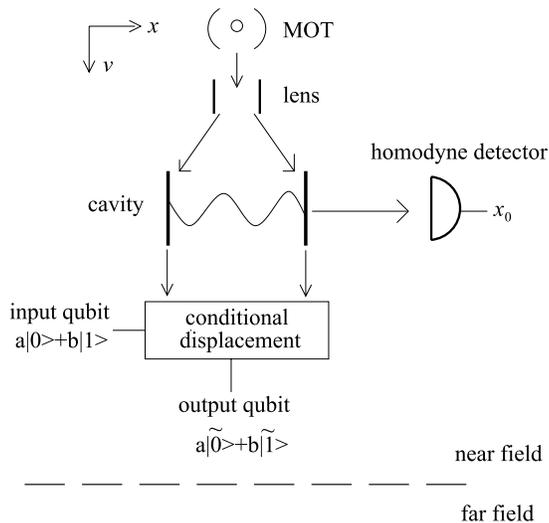}
\end{center}
\par
\vspace{-0.0cm}\caption{An atom cooled in a MOT freely falls
through a high finesse cavity, orthogonally to the cavity axis.
Above the cavity (and relatively far from it) a diverging atomic
lens causes a delocalization of the atomic wave-function entering
the cavity. The atom interacts with a single mode of the cavity
and after a suitable interaction time, the intracavity quadrature
$\hat{x}_{0}=\hat{c}+\hat{c}^{\dagger}$ is measured and the atomic
wave-function is projected onto an approximate comb-like state,
which we take as the approximate codeword $\widetilde
{|0\rangle}$. A conditional displacement (see text) can then be
used to generate any correctable state
$a\widetilde{|0\rangle}+b\widetilde{|1\rangle}
$.}%
\label{setup}%
\end{figure}

Our scheme concerns a two-level atom transversally crossing a high
finesse optical cavity and interacting with one of its modes (see
Fig.~\ref{setup} for a schematic description). We shall see that,
if at an appropriate interaction time a homodyne measurement of an
intracavity quadrature is performed, the center-of-mass motion of
the atom is projected onto an approximate comb-like state, which
can be used for the generation of the approximated codeword
states. Notice that here we are encoding a qubit into the external
degrees of freedom of a \emph{free} atom, which can be always seen
as a quantum oscillator with zero frequency.

This set-up can be realized using a small sample of atoms cooled in a
magneto-optical-trap (MOT) \cite{Metcalf} placed above the optical cavity. The
atoms are then let fallen down one by one through the cavity and if the MOT is
distant enough from the cavity, and using appropriate collimators, the atom
velocity is exactly orthogonal to the cavity axis $x$. We want to encode
qubits into the CV corresponding to the atomic motion along $x$ and the
relevant dynamics is described (assuming, as usual in the optical domain, the
rotating wave and dipole approximation) by the following Hamiltonian
\cite{walls}
\begin{equation}
\hat{H}=\hbar\omega_{0}\hat{\sigma}_{z}+\frac{\hat{p}^{2}}{2M}+\hbar\omega
_{c}\hat{c}^{\dagger}\hat{c}+\hbar g_{0}\left(  \hat{\sigma}^{\dagger}\hat
{c}+\hat{c}^{\dagger}\hat{\sigma}\right)  \cos k_{c}\hat{x}\,. \label{ham2}%
\end{equation}
In this Hamiltonian $\hat{\sigma}_{z}=(\hat{\sigma}^{\dagger}\hat{\sigma}%
-\hat{\sigma}\hat{\sigma}^{\dagger})/2$, $\hat{\sigma},\hat{\sigma}^{\dagger}$
are the atomic spin-$1/2$ operators associated with the two internal levels
whose transition is quasi-resonant with the optical cavity mode, $\hat{x}%
,\hat{p}$ are the atomic center-of-mass position and momentum operators along
$x$, $M$ is the atomic mass, $\omega_{0}=2\pi c/\lambda_{0}$ is the atomic
transition frequency ($c$ is the speed of light), $\hat{c},\hat{c}^{\dagger}$
are the cavity mode annihilation and creation operators, $\omega_{c}%
=ck_{c}=2\pi c/\lambda_{c}$ is the cavity mode frequency, and $g_{0}$ is the
atom-field coupling constant.

A ponderomotive interaction between the atom and the cavity mode is obtained
in the dispersive limit in which the cavity mode is highly (red) detuned from
the atomic transition. In this limit, the upper atomic level can be
adiabatically eliminated and also the spontaneous emission from it can be
neglected. In such a condition the atom always remains in its ground state and
the resulting ponderomotive Hamiltonian is (in a frame rotating at the
frequency $\omega_{c}$)
\begin{equation}
\hat{H}=\frac{\hat{p}^{2}}{2M}-\frac{\hbar g_{0}^{2}}{\delta}\hat{c}^{\dagger
}\hat{c}~\cos^{2}k_{c}\hat{x}, \label{ham2b}%
\end{equation}
where $\delta\equiv\omega_{0}-\omega_{c}$ is the detuning. We then make a
second assumption, the so called Raman-Nath approximation \cite{walls}, which
amounts to assume that the interaction time $t$ (given by the time the atom
takes to cross the cavity mode, i.e. $t\simeq2w_{0}/v$, where $w_{0}$ is the
cavity mode waist and $v$ is the atom velocity) is short enough so that any
variation of the atomic kinetic energy along $x$ due to photon exchanges with
the cavity field can be neglected. In this limit the kinetic energy along the
cavity axis becomes a constant of motion, equal to its value before the cavity
crossing, and therefore can be eliminated from the Hamiltonian of
Eq.~(\ref{ham2b}).

At the beginning the cavity mode is in a coherent state $\left|
\alpha\right\rangle _{c}$ (we can always choose the phase reference so that
$\alpha\geq0$), while the atomic motion along the cavity axis $x$ is described
by a generic wave-function $\Phi\left(  x\right)  $, so that the initial state
of the system is
\begin{equation}
\left|  \Psi\left(  0\right)  \right\rangle =\left|  \alpha\right\rangle
_{c}\otimes\int dx\Phi\left(  x\right)  \left|  x\right\rangle _{a}=\int
dx\Phi\left(  x\right)  \left|  \alpha,x\right\rangle \,. \label{condin}%
\end{equation}
At the end of the atom-cavity interaction, i.e. after the interaction time $t
$, the state of the system becomes \cite{walls}
\begin{equation}
\left|  \Psi\left(  t\right)  \right\rangle =e^{-\frac{i}{\hbar}\hat{H}%
t}\left|  \Psi\left(  0\right)  \right\rangle =\int dx\Phi\left(  x\right)
\left|  \alpha\left(  x,t\right)  ,x\right\rangle \,,
\end{equation}
where
\begin{equation}
\alpha\left(  x,t\right)  =\alpha_{1}\left(  x,t\right)  +i\alpha_{2}\left(
x,t\right)  =\alpha\exp\left(  i\frac{g_{0}^{2}t}{\delta}\cos^{2}%
k_{c}x\right)  \,. \label{alfatot}%
\end{equation}
Just at the end of the interaction we measure the intracavity quadrature
$\hat{x}_{0}=\hat{c}+\hat{c}^{\dagger}$ \cite{fast} obtaining the result
$x_{0}$. As a consequence the cavity mode is projected onto the corresponding
quadrature eigenstate $\left|  x_{0}\right\rangle $, while the atomic motion
along $x$ is disentangled from the cavity mode and it is projected onto the
state with wave-function \cite{walls}
\begin{equation}
\Phi_{x_{0}}\left(  x,t\right)  =N_{x_{0},t}\Phi\left(  x\right)  \exp\left\{
- \left[  \alpha_{1}\left(  x,t\right)  -\frac{x_{0}}{2}\right]  ^{2}%
-i\alpha_{2}\left(  x,t\right)  \left[  \alpha_{1}\left(  x,t\right)
-x_{0}\right]  \right\}  \,, \label{near_field}%
\end{equation}
where $N_{x_{0},t}$ is a normalization constant. Now it is possible to see
that this state becomes a comb-like state with well localized spikes, so that
it can be used as an approximate codeword, if we choose the interaction time
$t$ where we make the homodyne measurement such that $g_{0}^{2}t/\delta=\pi$,
and we take as initial wave-function $\Phi\left(  x\right)  $ a completely
delocalized state, i.e.
\begin{equation}
\Phi(x)=\left\{
\begin{array}
[c]{cl}%
L^{-\frac{1}{2}} & \quad0\leq~x\leq L\\
0 & \quad x<0,~x>L
\end{array}
\right.  \,, \label{unif}%
\end{equation}
which is an approximate momentum eigenstate with $p=0$ ($L$ is the
cavity length). Such a delocalized state can be prepared using a
suitable diverging atomic lens, i.e. an antinode of a blue-detuned
cavity or another repulsive quadratic optical potential
\cite{Metcalf,lente}, soon after the MOT and before the atom
enters the cavity (see Fig.~\ref{setup}). With the above choices,
the atomic wave-function of Eq.~(\ref{near_field}) takes the
following form
\begin{equation}
\Phi_{x_{0}}(x)=\left\{
\begin{array}
[c]{cl}%
N_{x_{0}}\exp\{-[\alpha_{1}(x)-\frac{x_{0}}{2}]^{2}-i\alpha_{2}(x)[\alpha
_{1}(x)-x_{0}]\} & \quad0\leq~x\leq L\\
0 & \quad x<0,~x>L
\end{array}
\right.  \label{fi}%
\end{equation}
where $\alpha_{1}\left(  x\right)  $ and $\alpha_{2}\left(  x\right)  $ are
given by Eq.~(\ref{alfatot}) with $g_{0}^{2}t/\delta=\pi$ and the
normalization constant $N_{x_{0}}$ is given by
\begin{equation}
N_{x_{0}}=\sqrt{\frac{k_{c}}{d~J\left(  \alpha,x_{0}\right)  }}\exp\left(
\frac{x_{0}^{2}}{4}\right)  \,, \label{N}%
\end{equation}
where $d=2L/\lambda_{c}$ is the integer number of half-wavelengths of the
stationary cavity mode and
\begin{equation}
J\left(  \alpha,x_{0}\right)  \equiv\int_{0}^{\pi}dy\exp\left\{  2\alpha
_{1}\left(  y\right)  \left[  x_{0}-\alpha_{1}\left(  y\right)  \right]
\right\}  \,. \label{J}%
\end{equation}
The normalization factor is connected with the probability density of the
outcome $x_{0}$ of the homodyne measurement, which is given by
\begin{equation}
\mathcal{P}\left(  x_{0}\right)  =\frac{J\left(  \alpha,x_{0}\right)  }%
{\sqrt{2\pi^{3}}}\exp\left(  -\frac{x_{0}^{2}}{2}\right)  \,. \label{P}%
\end{equation}

In order to make a direct comparison with the ideal codewords of
Sec.~\ref{ideal} and to simplify the formulas, in the following we
adopt dimensionless position and momentum operators by setting
$\lambda_{c}=\hbar=1$. It is also convenient to consider the
scaled dimensionless position variable $y=k_{c}x$, as we have
already done in Eq.~(\ref{J}) where $\alpha_{1}\left( y\right)
=\alpha_{1}\left( x=y/k_{c}\right)  $.

\subsection{Homodyning with a zero outcome}

\begin{figure}[ptbh]
\vspace{+0.1cm}
\par
\begin{center}
\includegraphics[width=0.75\textwidth] {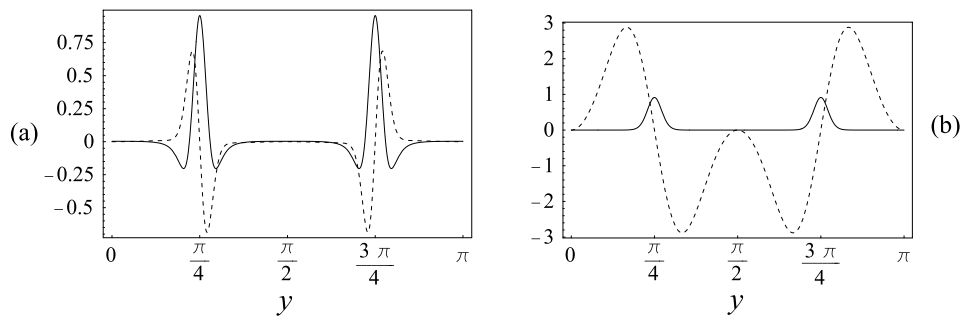}
\end{center}
\par
\vspace{-0.6cm}\caption{Approximate codeword $\widetilde{\left|
0\right\rangle } $ corresponding to the parameters $\alpha=2.4$ and $d=20$. In
(a) the real part of the spatial wave-function Re$\varphi_{0}$ (solid line)
and the imaginary part Im$\varphi_{0}$ (dashed line) are plotted vs the scaled
position variable $y$ inside a single period. In (b) the spatial probability
distribution $\left|  \varphi_{0}\right|  ^{2} $ (solid line) and the phase
$-i\ln[\varphi_{0}/\left|  \varphi_{0}\right|  ]$ (dashed line) are plotted vs
the scaled position variable $y$ inside a single period. }%
\label{fig02b}%
\end{figure}

\begin{figure}[ptbh]
\vspace{-0.0cm}
\par
\begin{center}
\includegraphics[width=0.8\textwidth] {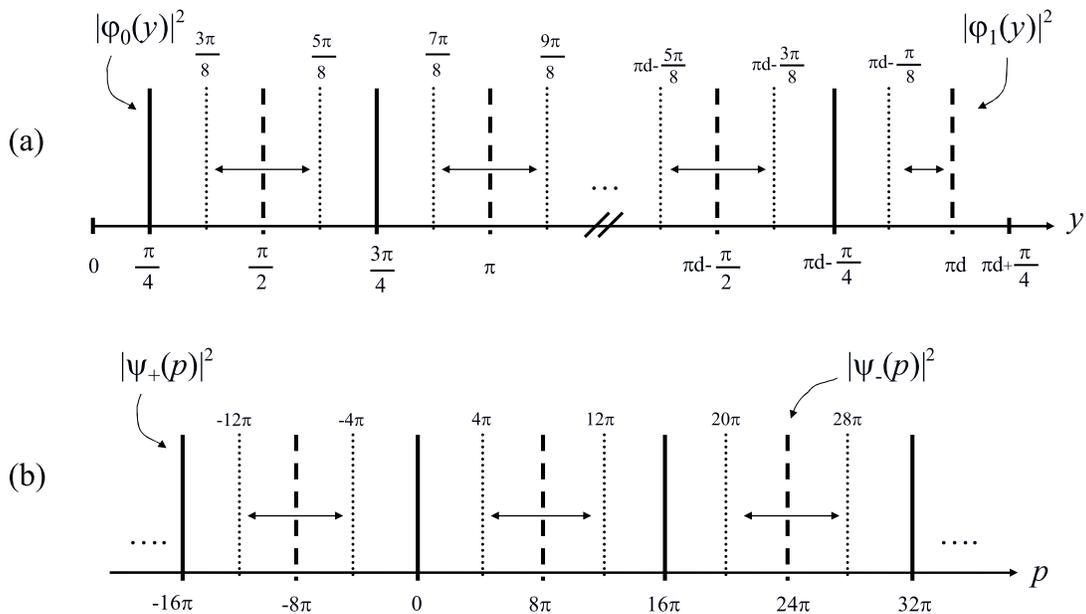}
\end{center}
\par
\vspace{-0.3cm}\caption{\textbf{(a)} Structure of the spatial probability
distributions $\left|  \varphi_{0}\right|  ^{2}$ (solid lines), $\left|
\varphi_{1}\right|  ^{2}$ (dashed lines)\ of the approximate codewords
$\widetilde{\left|  0\right\rangle },\widetilde{\left|  1\right\rangle }$ vs
the scaled variable $y$. The two distributions are displaced by $\pi/4$ and
each of them has $\pi/2$-spaced spikes. Dotted lines and arrows delimit the
error regions $R_{n},R_{2d}$ defined in the text. \textbf{(b)} Structure of
the momentum probability distributions $\left|  \psi_{+}(p)\right|  ^{2}$
(solid lines), $\left|  \psi_{-}(p)\right|  ^{2}$ (dashed lines) of the
approximate codewords $\widetilde{\left|  +\right\rangle },\widetilde{\left|
-\right\rangle }$. The two distributions are displaced by $8\pi$ and each of
them has $16\pi$-spaced spikes. Dotted lines and arrows delimit the error
regions $R_{n}^{+}$ defined in the text.}%
\label{fig03}%
\end{figure}

For the sake of simplicity, we shall consider from now on the particular case
of a homodyne measurement result, $x_{0}=0$. First of all we define the atomic
state of Eq.~(\ref{fi}) with $x_{0}=0$, as the approximate codeword
$\widetilde{\left|  0\right\rangle }$, i.e.
\begin{equation}
\varphi_{0}\left(  x\right)  \equiv\left\langle x\right.  \widetilde{\left|
0\right\rangle }=\Phi_{x_{0}=0}\left(  x\right)  =N_{0}\exp[-\alpha_{1}\left(
x\right)  ^{2}-i\alpha_{1}\left(  x\right)  \alpha_{2}\left(  x\right)
]\,,\text{ \ \ \ \ \ \ }\left(  0\leq x\leq L\right)  \label{fizero}%
\end{equation}
where $N_{0}\equiv N_{x_{0}=0}$. One can verify that the resulting
wave-function $\varphi_{0}\left(  y\right)  $ is periodic in $0\leq y\leq\pi
d$ with period equal to $\pi$ and it has $2d$ equally-spaced spikes (i.e. with
$\pi/2$-spacing, see Fig.~\ref{fig02b}(a)), so that its choice as approximate
codeword state $\widetilde{\left|  0\right\rangle }$ is justified. From such a
state it is easy to generate the associated codeword state $\widetilde{\left|
1\right\rangle }$ by simply displacing in $y$ the state $\widetilde{\left|
0\right\rangle }$ by the quantity $\pi/4$, so that the corresponding
wave-function is $\varphi_{1}\left(  y\right)  =\varphi_{0}\left(
y-\pi/4\right)  $ (which is nonzero in $\pi/4\leq y\leq\pi d+\pi/4$, see
Fig.~\ref{fig03} for a schematic description of the corresponding probability
distributions). The practical implementation of the displacement of the atomic
wave-function can be achieved by applying, just after the cavity, a suitable
electric field gradient pulse with an appropriate intensity.

After seeing how the two basis codeword states are generated, let us now see
how to generate a generic superposition of the two, $a\widetilde{\left|
0\right\rangle }+b\widetilde{\left|  1\right\rangle }$. These superpositions
can be generated using conditional displacement schemes analogous to those
used, for example, in the manipulation of quantum states of trapped ions
\cite{wineRMP} and which exploit the coupling of a motional degree of freedom
with an internal transition of the ion. Schematically these schemes proceed as
follows. The atom is prepared in the tensor product state $\widetilde{\left|
0\right\rangle }\otimes\left[  a|g\rangle+b|e\rangle\right]  $, where
$|e\rangle$ and $|g\rangle$ are two ground state sublevels. Then a laser pulse
which is only coupled to $|e\rangle$ is applied to the atom and its intensity
is tuned so to give exactly a position shift $y\rightarrow y-\pi/4$. In this
way the state of the atom becomes $a|g\rangle\otimes\widetilde{\left|
0\right\rangle }+b|e\rangle\otimes\widetilde{\left|  1\right\rangle }$. Then a
\emph{rf pulse} resonant with the $e\rightarrow g$ transition and transforming
$|e\rangle\rightarrow(|e\rangle+|g\rangle)/\sqrt{2}$ and $|g\rangle
\rightarrow(|g\rangle-|e\rangle)/\sqrt{2}$ is applied, so that the state of
the atom becomes $[|g\rangle\otimes(a\widetilde{\left|  0\right\rangle
}+b\widetilde{\left|  1\right\rangle })+|e\rangle\otimes(b\widetilde{\left|
1\right\rangle }-a\widetilde{\left|  0\right\rangle })]/\sqrt{2}$. When the
internal state of the atom is measured and it is found equal to $|g\rangle$,
the atomic motional state is conditionally generated in the desired encoded
superposition $a\widetilde{\left|  0\right\rangle }+b\widetilde{\left|
1\right\rangle }$. Examples of superposition states are also the approximate
eigenstates $\widetilde{\left|  \pm\right\rangle }$ of the phase-flip operator
$\bar{X}$ and equivalent set of codewords, which are given by $\widetilde
{\left|  \pm\right\rangle }\equiv\lbrack\widetilde{\left|  0\right\rangle }%
\pm\widetilde{\left|  1\right\rangle }]/{\mathcal{N}_{\pm}}$, where
$\mathcal{N}_{\pm}^{2}=2(1\pm\mathrm{Re}[\widetilde{\left\langle 0\right.
}\widetilde{\left|  1\right\rangle }])$ because $\widetilde{\left|
0\right\rangle }$ and $\widetilde{\left|  1\right\rangle }$ are not exactly
orthogonal in general. Their wave-function $\varphi_{\pm}\left(  y\right)
=\left[  \varphi_{0}\left(  y\right)  \pm\varphi_{1}\left(  y\right)  \right]
/\mathcal{N}_{\pm}$ are nonzero in $0\leq y\leq\pi d+$ $\pi/4$ and have spikes
spaced by $\pi/4$. However these approximated codewords have to be close the
ideal ones also in momentum space. Performing the Fourier transform of the
above wave-functions, it is possible to see that the momentum wave-function of
$\widetilde{\left|  0\right\rangle }$ and $\widetilde{\left|  1\right\rangle
}$, $\psi_{0}\left(  p\right)  $ and $\psi_{1}\left(  p\right)  $, have
equally spaced spikes separated by $8\pi$, which coincide for even $n$ and are
opposite for odd $n$, due to the relation $\psi_{0}\left(  p\right)
=e^{ip/8}\psi_{1}\left(  p\right)  $, which is an immediate consequence of the
translation by $\pi/4$ in the position coordinate. As a consequence,
$\ \psi_{\pm}\left(  p\right)  =\left[  \psi_{0}\left(  p\right)  \pm\psi
_{1}\left(  p\right)  \right]  /\mathcal{N}_{\pm}$ have spikes spaced by
$16\pi$ and shifted by $8\pi$ with respect to each other (see Fig.~\ref{fig03}%
~(b) for a schematic representation of the probability
distributions in momentum space). Therefore, from these
considerations, and comparing Fig.~\ref{fig03} with the
description of the ideal codewords states in Fig.~\ref{fig00}, we
can conclude that the states generated in this lithographic scheme
can certainly be used as approximated codeword states in
the case of a spacing parameter $\theta=1/8$ (see Eqs.(\ref{ideal_0}%
)-(\ref{ideal_meno})). In such a case in fact, the structure of the peaks is
recovered both in position and momentum space for the four codewords, even
though, as expected, the approximated codewords have a finite number of peaks
$(2d)$ and the peaks have a nonzero width and a finite height.

It is important to notice that, unfortunately, the codeword states
generated in this way can be used only when the atoms are not too
far from the cavity (\emph{near field} regime). In fact, after
leaving the cavity, the atomic motion along $x$ evolves as a free
particle and this evolution leads to quantum interference between
the various spikes (see Ref.~ \cite{walls}). As we can see from
Fig.~\ref{fig02b}(b), the phase change of the front of the atomic
wave-function is approximately linear at the position $y$ where
$\left| \varphi_{0}\left( y\right)  \right| ^{2}\neq0$; for this
reason, the various spikes are deflected after the cavity and they
interfere in the \emph{far field} \cite{walls}.

\subsection{Intrinsic error probability \label{error_recovery}}

As discussed in Sec.~\ref{ideal}, when approximated codewords are used, one
has additional errors (intrinsic errors). In fact, due to the presence of the
tails of the peaks, the recovery process may lead sometimes to a wrong
codeword. The recovery in the spatial variable is performed by measuring the
operator $\hat{y}($mod$\pi/4)$. We can see from Fig.~\ref{fig03}~(a) that an
intrinsic error in the recovery process occurs when, given the state
$\varphi_{0}\left(  y\right)  $, the measurement gives a result within one of
the error regions: $R_{n}\equiv\left[  (4n-1)(\pi/8),(4n+1)(\pi/8)\right]  $,
$n=1,...,2d-1$ and $R_{2d}\equiv\left[  \pi d-\pi/8,\pi d\right]  $. In such a
case, in fact, the original state $\widetilde{\left|  0\right\rangle }$ will
be correct to the other one $\widetilde{\left|  1\right\rangle }$ corrupting
the encoded information even in the absence of any errors of the channel. The
corresponding error probability $P_{x,0}$ is equal to the one, $P_{x,1}$,
which we would obtain starting from the state $\varphi_{1}\left(  y\right)  $
and considering the complementary error region $\left[  \pi/4,\pi d\right]
\backslash\cup_{n=1,2d}R_{n}$. So, we simply have
\begin{equation}
P_{x}=\sum_{n=1}^{2d}\int_{R_{n}}\frac{dy}{2\pi}\left|  \varphi_{0}\left(
y\right)  \right|  ^{2}=\frac{\left(  4d-1\right)  N_{0}^{2}}{2\pi}\int
_{0}^{\frac{\pi}{8}}dy\exp\left[  -2\alpha_{1}\left(  y\right)  ^{2}\right]
\,. \label{Px}%
\end{equation}
The recovery in momentum space is done by measuring the operator $\hat{p}%
($mod$8\pi)$. In the same way, one can define (see Fig.~\ref{fig03}~(b)) the
two different error regions: $R_{n}^{+}\equiv\left[  \left(  2n+1\right)
8\pi-4\pi,\left(  2n+1\right)  8\pi+4\pi\right]  $ and $R_{n}^{-}\equiv\left[
(2n)8\pi-4\pi,(2n)8\pi+4\pi\right]  $ with $n=0,\pm1,...$ An error in the
recovery process occurs when, given the state $\ \psi_{\pm}\left(  p\right)
$, the measurement gives a result within one of the error regions $R_{n}^{\pm
}$. The corresponding error probability is then given by
\begin{equation}
P_{p,\pm}=\sum_{n=-\infty}^{+\infty}\int_{R_{n}^{\pm}}dp\left|  \psi_{\pm
}\left(  p\right)  \right|  ^{2}=\frac{2}{\mathcal{N}_{\pm}^{2}}%
\sum_{n=-\infty}^{+\infty}\int_{R_{n}^{\pm}}dp\left(  1\pm\cos\frac{p}%
{8}\right)  \left|  \psi_{0}\left(  p\right)  \right|  ^{2}\,. \label{probp}%
\end{equation}
Exploiting the parity of $\left|  \psi_{0}\left(  p\right)  \right|  ^{2}$ and
the inequality (true almost everywhere)
\begin{equation}
\left|  \psi_{0}\left(  p\right)  \right|  ^{2}\leq\frac{4N_{0}^{2}}{\pi}%
\frac{\sin^{2}(\frac{pL}{2})}{p^{2}}\,, \label{maggio}%
\end{equation}
we obtain
\begin{equation}
P_{p,+}\leq\frac{16}{\pi}\frac{N_{0}^{2}}{\mathcal{N}_{+}^{2}}\sum
_{n=0}^{+\infty}\int_{(4n+1)4\pi}^{(4n+3)4\pi}dp\left(  1+\cos\frac{p}%
{8}\right)  \frac{\sin^{2}(pL/2)}{p^{2}}\equiv P_{+}\,, \label{Ppiu}%
\end{equation}
and
\begin{equation}
P_{p,-}\leq\frac{8}{\pi}\frac{N_{0}^{2}}{\mathcal{N}_{-}^{2}}\left\{
\int_{-4\pi}^{4\pi}dp\left(  1-\cos\frac{p}{8}\right)  \frac{\sin^{2}%
(pL/2)}{p^{2}}+2\sum_{n=1}^{+\infty}\int_{(4n-1)4\pi}^{(4n+1)4\pi}dp\left(
1-\cos\frac{p}{8}\right)  \frac{\sin^{2}(pL/2)}{p^{2}}\right\}  \equiv
P_{-}\,. \label{Pmen}%
\end{equation}
To estimate the quality of the overall encoding procedure provided by our
scheme, we have to consider a mean intrinsic error probability $\bar{P}_{e}$,
which is obtained in general by averaging over all the possible encoded qubit
states. Using the above definitions, we have that the mean intrinsic error
probability $\bar{P}_{e}$ satisfies the inequality
\begin{equation}
\bar{P}_{e}\lesssim\max\left\{  P_{x},P_{p,+},P_{p,-}\right\}  \leq
\max\left\{  P_{x},P_{+},P_{-}\right\}  \equiv P_{\max}\,, \label{pmax}%
\end{equation}
which defines the maximum intrinsic error probability $P_{\max}$, providing
therefore a good characterization of the proposed encoding scheme.

We have therefore to estimate $P_{\max}$ in the case of implementation on a
realistic cavity QED apparatus (see for example Ref.~\cite{kimble}). In
general the error probabilities $P_{x},P_{p,\pm}$ depend on two dimensionless
parameters: $\alpha$, the amplitude of the initial coherent field in the
cavity, and $d$, which is connected to the cavity length. These parameters
cannot be taken at will however, because we have to satisfy the assumptions
used for the derivation of the approximated codeword states, namely the large
detuning and the Raman-Nath approximations. This latter approximation can also
be expressed by imposing that the uncertainty of the position along the cavity
axis acquired by the atom during the interaction time is much smaller than the
cavity mode wavelength, i.e. $\Delta x\ll\lambda_{c}$ \cite{walls} (here, we
re-introduce physical dimensions, in order to fully describe the experimental
implementation). One has $\Delta p\simeq\alpha^{2}\hbar k_{c}$ \cite{walls},
from which we get $\Delta x\simeq\Delta p~t/2M\simeq\alpha^{2}\hbar k_{c}%
t/2M$, so that the Raman-Nath approximation implies the following condition on
the interaction time
\begin{equation}
t\ll\frac{M\lambda_{c}^{2}}{\pi\hbar\alpha^{2}}\,. \label{tRN}%
\end{equation}
On the other hand, the condition of large detuning implies $4\alpha^{2}%
g_{0}^{2}/\delta^{2}\ll1$ which, together with the condition $g_{0}%
^{2}t/\delta=\pi$ used above, leads to another condition on the interaction
time, i.e.
\begin{equation}
t\gtrsim\frac{2\pi\alpha}{g_{0}}, \label{largedet}%
\end{equation}
which, combined with Eq.~(\ref{tRN}), gives the following bounds for the
interaction time
\begin{equation}
\frac{2\pi\alpha}{g_{0}}\lesssim t\ll\frac{M\lambda_{c}^{2}}{\pi\hbar
\alpha^{2}}. \label{largedetfin}%
\end{equation}
This condition however puts also limitations on the possible values of
$\alpha$ and of the coupling constant $g_{0}$, which is related to the cavity
mode volume $V$ (and therefore to $d$ because it is $V\sim\pi w_{0}^{2}L=\pi
w_{0}^{2}d\lambda_{c}/2$) by the relation $g_{0}=d_{12}\sqrt{\omega_{c}%
/2\hbar\varepsilon_{0}V}$, where $d_{12}$ is the electric dipole matrix
element associated to the atomic transition and $\varepsilon_{0}$ is the
vacuum dielectric constant. In order to satisfy Eq.~(\ref{largedetfin}) one
can impose, for example, $2\pi\alpha/g_{0}=10^{-2}M\lambda_{c}^{2}/\pi
\hbar\alpha^{2}$ which becomes therefore an effective relation between the two
apparently independent parameters $\alpha$ and $d$, which reads
\begin{equation}
\alpha=\sqrt[3]{\frac{g_{0}}{D}}\text{ , \ }D\equiv10^{2}\frac{2\pi^{2}\hbar
}{M\lambda_{c}^{2}}\,. \label{alfaD}%
\end{equation}
To state it in other words, the assumptions made in order to derive the
desired encoded states implies that in practice we have only \emph{one} free
parameter, which can be $\alpha$, $d$ or the coupling constant $g_{0}$.

\begin{figure}[ptbh]
\vspace{+0.2cm}
\par
\begin{center}
\includegraphics[width=0.5\textwidth] {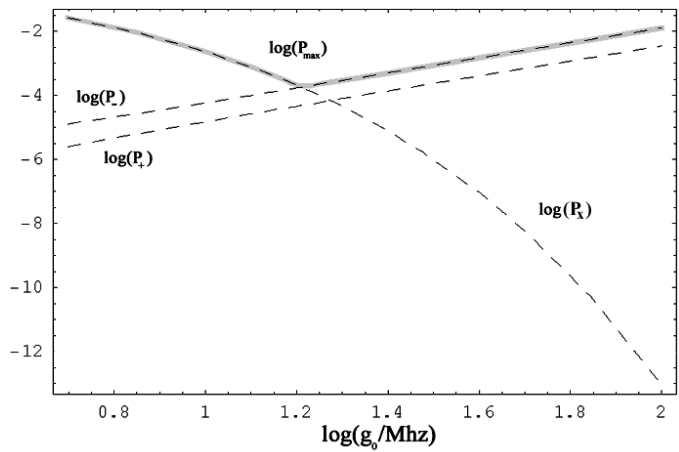}
\end{center}
\par
\vspace{-0.6cm}\caption{$\log_{10}(P_{x})$ and $\log_{10}(P_{\pm})$
(dashed-lines) versus $\log_{10}(g_{0}/$MHz$)$ in the case of $Cs$ and for a
cavity waist $w_{0}=20\mu$m. The quantity $\log_{10}(P_{max})$ is the marked
upper curve and it displays a minimum at about $g_{0}=16$MHz. In
correspondence of such a minimum we have a mean error probability $\bar{P}%
_{e}\lesssim P_{\max}\sim2\times10^{-4}$. }%
\label{fig05}%
\end{figure}

To show the experimental feasibility of the present scheme, we
have considered the case of an heavy atom ($Cs$) and we have
studied the behavior of the error probabilities $P_{x},P_{\pm}$ in
the case of realistic parameters. In particular, we have
considered $\lambda_{0}=\lambda_{c}=852.1$ nm and
$d_{12}=3.79\times10^{-29}$ Cm, so that $D\simeq1.3\times10^{6}$
Hz. In Fig.~\ref{fig05} we have plotted the three error
probabilities and the corresponding maximum probability $P_{max}$
as a function of the coupling constant $g_{0}$ in the case of a
cavity mode waist $w_{0}=20\mu$m. We can see that the error
probabilities in position and in momentum behave in the opposite
way for increasing $g_{0}$ and, for this reason, the upper bound
$P_{\max}$ has a minimum at an intermediate value $g_{0}\simeq16$
MHz, where all the probabilities $P_{x},P_{\pm}$ have about the
same order of magnitude, i.e. $\sim10^{-4}$, which represents a
remarkably small value of the intrinsic error probability. In such
a case, we have $\alpha\simeq2.3$, which gives
$\mathcal{P}(x_{0}=0)\simeq4.6\%$, while the interaction time is
$t\simeq3\mu $s, and the atom velocity is $v=2w_{0}/t\simeq40$
ms$^{-1}$.

\section{Conclusion\label{conclusion}}

Continuous variable QEC consists in encoding quantum information
(i.e. a qubit) into a quantum system whose state is described by
observables with a continuous spectrum of eigenvalues. The
redundancy of these CV degrees of freedom can be used to correct
the errors which arise from the unwanted interactions with the
environment, and therefore to safely protect the encoded quantum
information. However, there is a\emph{ fault} in the CV QEC theory
that concerns the physical generation of the CV\ quantum codewords
which, ideally, are non-normalizable states. In other words, every
real experimental setup can only make use of an approximate
version of such codewords, and it comes out the problem of how one
can generate such codewords and what are the consequent effects in
terms of error correcting performances. Here, to face the problem,
we have resorted to lithographic techniques. In particular, we
have shown how an optical cavity subjected to a homodyne
measurement acts as a virtual diffraction grating which is able to
project the motional state of a crossing neutral atom into a
well-approximated quantum codeword. Actually, this CV encoding is
limited in space, i.e., the generated CV quantum codewords will
live only in the near-field regime, since they will be destroyed
in the far-field regime due to quantum diffraction. However, under
these assumptions, we have shown that sufficiently low values of
the intrinsic error probability are effectively reachable (in
particular $\sim10^{-4}$ using a Cesium atom).


\begin{thebibliography}{99}
\bibitem{cvbook}See e.g. \textit{Quantum Information Theory with Continuous
Variables}, edited by A. K. Pati and S. L. Braunstein, Kluwer Academic Press (2002).

\bibitem {Lloyd2}S. Braunstein, Phys. Rev. Lett. \textbf{80}, 4084 (1998); S.
Lloyd and J. E. Slotine, Phys. Rev. Lett. \textbf{80}, 4088 (1998).

\bibitem {preskill}D. Gottesman, A. Kitaev, and J. Preskill, Phys. Rev. A
\textbf{64}, 012310 (2001).

\bibitem {Travaglione0}B. C. Travaglione and G. J. Milburn, Phys. Rev. A
\textbf{66}, 052322 (2002).

\bibitem {Travaglione}B. C. Travaglione and G. J. Milburn, Phys. Rev. A
\textbf{65}, 032310 (2002).

\bibitem {epl}S. Pirandola, S. Mancini, D. Vitali and P. Tombesi, Europhys.
Lett. \textbf{68}, 323 (2004).

\bibitem {ponde}S. Pirandola, S. Mancini, D. Vitali and P. Tombesi, quant-ph/0503003 (accepted for publication on EPJD).

\bibitem {Giovannetti}S. Mancini and P. Tombesi, Phys. Rev. A \textbf{49},
4055 (1994).

\bibitem {walls}P. Storey, M. Collett, and D. F. Walls, Phys. Rev. Lett.
\textbf{68}, 472 (1992).

\bibitem {wallsB}D. F. Walls, Aust. J. Phys. \textbf{49}, 715 (1996).

\bibitem {particle}A one-dimensional quantum oscillator or a one-dimensional
free particle, which is a $1-\dim$ quantum oscillator with zero frequency.

\bibitem {stabilizer}D. Gottesman, Phys. Rev. A \textbf{54}, 1862 (1996); A.
R. Calderbank, E. M. Rains, P. W. Shor, and N. J. A. Sloane, Phys. Rev. Lett.
\textbf{78}, 405 (1997).

\bibitem {Metcalf}H. J. Metcalf and P. van der Straten, \textit{Laser Cooling
and Trapping} (Springer, 1999).

\bibitem {fast}It is possible to make a direct and precise measurement of this
intracavity quantity using a high finesse cavity whose input-output mirror
transmittivity is controlled through fast electronics \cite{Tau}.

\bibitem {Tau}M. S. Taubman, H. M. Wiseman, D. E. McClelland, and H. A.
Bachor, J. Opt. Soc. Am. B \textbf{12}, 1792 (1995).

\bibitem {lente}J. Bjorkholm, R. Freeman, A. Ashkin, and D. Pearson, Phys.
Rev. Lett. \textbf{41}, 1361 (1978).

\bibitem {wineRMP}D. Leibfried, R. Blatt, C. Monroe, and D. Wineland Rev. Mod.
Phys. \textbf{75}, 281 (2003).

\bibitem {kimble}J. McKeever, J. R. Buck, A. D. Boozer, A. Kuzmich, H.-C.
Nagerl, D. M. Stamper-Kurn, and H. J. Kimble Phys. Rev. Lett. \textbf{90},
133602 (2003).
\end{thebibliography}
\end{document}